\documentstyle{l-aa}    % formato AA
\def\ltsim{\raise 2pt \hbox {$<$} \kern-1.1em \lower 4pt \hbox {$\sim$}}
\def\ltapprox{\raise 2pt \hbox {$<$} \kern-1.1em \lower 5pt \hbox {$\approx$}}
\def\gtsim{\raise 2pt \hbox {$>$} \kern-1.1em \lower 4pt \hbox {$\sim$}}
\def\gtapprox{\raise 2pt \hbox {$>$} \kern-1.1em \lower 5pt \hbox {$\approx$}}

\def\skuno{\vskip 20pt}

\begin{document}
\thesaurus{02.16.2; 11.03.4 Coma; 11.09.1 NGC 4869; 11.09.3; 13.18.1}
\title
{ The magnetic field in the Coma cluster}
\skuno
\skuno
\author{L. Feretti\inst{1,2} \and D. Dallacasa\inst{3} \and
G. Giovannini\inst{1,2} \and A. Tagliani\inst{2}}

\offprints{L. Feretti}
\institute{
Dipartimento di Astronomia dell'Universit\'a,
Via Zamboni 33, I--40216 Bologna, Italy.
\and Istituto di Radioastronomia -- CNR, via Gobetti 101, I--40129
Bologna, Italy.
\and Joint Institute for VLBI in Europe, Postbus 2,
NL--7990 AA Dwingeloo, The Netherlands}

\maketitle

\begin{abstract}

The polarization data of the radio galaxy NGC4869, belonging to the Coma
cluster and located in its central region, allow us to obtain information on
the structure of magnetic field associated with the cluster itself. A magnetic
field of $\sim$ 8.5 $\mu$G, tangled on scales of the order of less than 1 kpc,
is required to explain the observed fluctuations of the rotation measure. This
magnetic field is more than one order of magnitude stronger than the
equipartition value obtained for Coma C. This implies that the halo source Coma
C may not be at the equipartition. Moreover, the need of efficient
reacceleration mechanisms for the electrons radiating in Coma C is stronger.
The energy supply to the Coma C radiating electrons is probably provided by the
cluster merger process.

\end{abstract}
\section{Introduction}

The existence of magnetic field associated with the intergalactic medium in
clusters of galaxies is indicated by the diffuse non thermal radio emission
(radio halos) revealed in some clusters. It has been independently suggested by
the existence of excess Faraday rotation measure of polarized radio emission in
radio sources within or behind the cluster (Kim et al. 1991) and from the
detection of Inverse Compton hard X-Ray emission (Bazzano et al. 1990). The
structure of this magnetic field is presently poorly known. Jaffe (1980)
suggested that it has to be tangled on a typical  galaxy size, while
Crusius-W\"atzel et al. (1990), studying the  depolarization in 5 strong double
sources,  find tangling on a smaller scale (0.5-2 kpc\footnote {A Hubble
constant $H_0$~=100~km~s$^{-1}$~Mpc$^{-1}$ is used throughout the paper, which
corresponds to a scale of 0.34 kpc/arcsec at the Coma cluster distance}). For
the Coma cluster of galaxies, Kim et al. (1990) give scale sizes for the
magnetic field reversals of 6.5-20 kpc.

The properties of the Coma cluster magnetic field, can be investigated through
the study of the extended radiogalaxy NGC4869 (5C4.81), located near the
cluster center, completely embedded within the halo source Coma C. This radio
source belongs to the class of Narrow Angle Tail (NAT) radio sources, which
represent the most extreme example of the interaction between a radio source
and the intracluster medium. A detailed study of this radio galaxy (an
elliptical with photographic magnitude 14.9$^m$  at $z=0.0232$) was performed
by us with the VLA, with multiple frequencies and resolutions (Dallacasa et al.
1989; Feretti et al. 1990). The radio structure consists of a faint flat
spectrum core, from which two straight short jets originate. They are sharply
bent at $\sim$~1~kpc  from the nucleus and produce two tails wrapping each
other and probably merging. This structure develops in EW for $\sim$~4
$^{\prime}$ (75 kpc) with some wide oscillations. Then the tail fades
progressively and shows a sharp bend to the North, which is detected only at
lower frequencies (Dallacasa et al. 1989). The tail spectrum steepens
progressively with the distance from the radio core, and at the source
periphery it is considerably steeper ($\alpha \sim 2.0$, with S$\sim
\nu^{-\alpha}$) than that of the surrounding regions of the radio halo Coma C
(Giovannini et al. 1993).

In this paper we present a study of the polarization properties of this galaxy
at 4 different wavelengths. We analyze the trends of the depolarization and
rotation measure in the jets and along the first arcmin of the tail, and
outline the properties of the depolarizing screen external to the source, i.e.
the Coma cluster.

We present the observational details in \S 2 and the images at the variuos
wavelenghts in \S 3. The multifrequency polarization comparison is given in \S
4, while in \S 5 we discuss the scenario emerging from the present results.

\section{Observations and data reduction}

Observations at 3.6, 6, 18 and 20.5 cm were performed with the Very Large Array
(VLA) in 1990 and 1991, in different configurations, with a bandwith of 50 MHz.
 The pointing was at RA=12$^h$56$^m$58.2$^s$,
DEC=28$^{\circ}$10$^{\prime}$50$^{\prime \prime}$, corresponding to the optical
and radio nucleus of NGC 4869. In Table 1 we give a summary of the
observational parameters.

\begin{table*}
\caption{Logs of observations and image parameters; $\sigma_I$ is the
rms noise in the total intensity image; $\sigma_{UQ}$ is the rms noise
in the Stokes' U and Q images.}
\begin{flushleft}
\begin{tabular}{cccccccc}
\hline
\noalign{\smallskip}
{}~Frequency~&$\lambda$ & Date & VLA Config. & Obs. Time &HPBW& $\sigma_I$
& $\sigma_{UQ}$\\
(MHz) & (cm) &  &    &(hours)&(arcsec)&(mJy/beam)& (mJy/beam) \\
\noalign{\smallskip}
\hline
\noalign{\smallskip}
 8415  & 3.6&  Nov 90   &  C    & 7.1  & 2.5 &    0.015   & 0.012  \\
       &    &  Apr 91   &  D    & \ \ 8.7 {\raise 4pt \hbox {$\}$}}  & & & \\
       &  &  &  &  &  &  &  \\
 4835  &  6 &  Nov 90   &  C    & 6.7  & 2.5 &    0.023   & 0.017  \\
       &    &  Apr 91   &  D    & \ \ 6.9  {\raise 4pt \hbox {$\}$}}  & & & \\
       &  &  &  &  &  &  &  \\
 1665  & 18 &  Nov 91   &  B    & 8.4  &  3  &    0.047   & 0.038  \\
       &    &  Nov 90   &  C    & \ \ 3.5  {\raise 4pt \hbox {$\}$}}  & & & \\
       &  &  &  &  &  &  &  \\
 1465  &20.5&  Nov 91   &  B    & 8.4  &  3  &    0.030   & 0.027  \\
       &    &  Nov 90   &  C    & \ \ 7.0  {\raise 4pt \hbox {$\}$}}  & & & \\
\noalign{\smallskip}
\hline
\noalign{\smallskip}
\end{tabular}
\end{flushleft}
\end{table*}

The flux-density scale and polarization position angle were calibrated by
observing 3C286. The scale of Baars et al. (1977) was assumed for the flux
density and the integrated polarization position angle of 3C286 was assumed
33$^{\circ}$ at all frequencies. The on-axis instrumental polarization of the
antennas was corrected using secondary calibrators observed with a wide range
of parallactic angles. The ionosferic Faraday rotation was corrected assuming
that the ionosphere is modelled by a thin sheet at the height of 350 km, with a
total electron content as provided by the monitoring station at Boulder
(Colorado). The phase calibration was made relative to the secondary
calibrators 1308+326 and 1345+125, after editing of bad datapoints.

Images in all Stokes parameters were produced with the NRAO AIPS package
following the standard procedure (Fourier inversion, Clean and Restore).
Self-calibration (Schwab 1980) was applied to minimise the effects of amplitude
and phase uncertainties of atmospheric and instrumental origin. The (u,v) data
at the same frequency but from different configurations were first reduced
separately, for a consistency check, then added together. At 6 cm, in order to
enhance the sensitivity and improve the uv coverage, we used also previous A, B
and D array data (Dallacasa et al. 1989).

The images in polarized intensity were obtained as P=(U$^2$+Q$^2$)$^{1/2}$, and
corrected for the positive bias arising from the previous combination. The
polarization angle was derived according to $\theta$=0.5 atan(U/Q).

The images at 3.6 and 6 cm were produced with the same uv range and restored
with an identical beam of 2.5$^{\prime\prime}$(HPBW). The images at 18 and 20.5
cm were made with HPBW=3$^{\prime\prime}$. The parameters relative to the
images (resolution and r.m.s noise) are given in Table 1.

\section{Results}

The full resolution images are presented in Figs 1 to 4. In all cases the lines
indicate the orientation of the electric vector, and their length is
proportional to the polarization percentage.

\begin{figure}
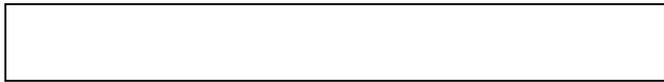
%*  dentro graffa per figura larga tutta la pagina
%%\special{psfile=POL6.PS.PS hoffset=-190 voffset=-535}
\picplace{1.0cm}    %   spazio che lascia per la figura
\caption{ Polarization image of NGC 4869 at 8.44 GHz. The restoring beam is a
circular Gaussian of HPBW = 2.5$^{\prime \prime}$.
Contour levels of the total intensity
image are --0.025,0.025,0.05,0.1,0.2,0.4,0.8,1.5,2.5 mJy/beam.
The E-vector length is proportional to the polarization
percentage: 1 arcsec corresponds to 5\%.}
\end{figure}%*  dentro graffa per figura larga tutta la pagina

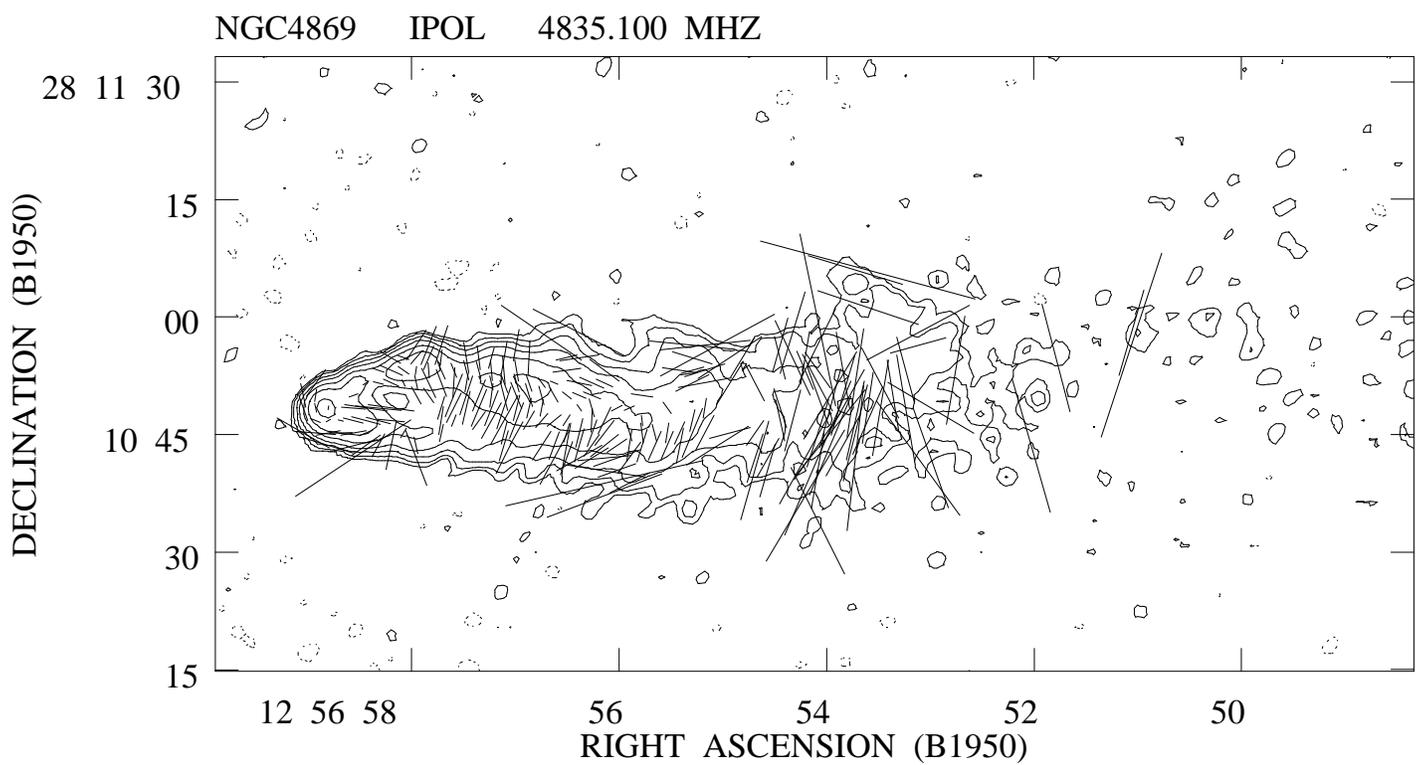
\begin{figure}
%%\special{psfile=POL6.PS.PS hoffset=-190 voffset=-535}
\picplace{1.0cm}
\caption{ Polarization image of NGC 4869 at 4.84 GHz. The restoring beam is a
circular Gaussian of HPBW = 2.5$^{\prime \prime}$. Contour
levels of the total intensity
image are --0.04,0.04,0.08,0.15,0.3,0.6,1,2 mJy/beam.
The E-vector length is proportional to the polarization
percentage: 1 arcsec corresponds to 6.25\%.}
\end{figure}

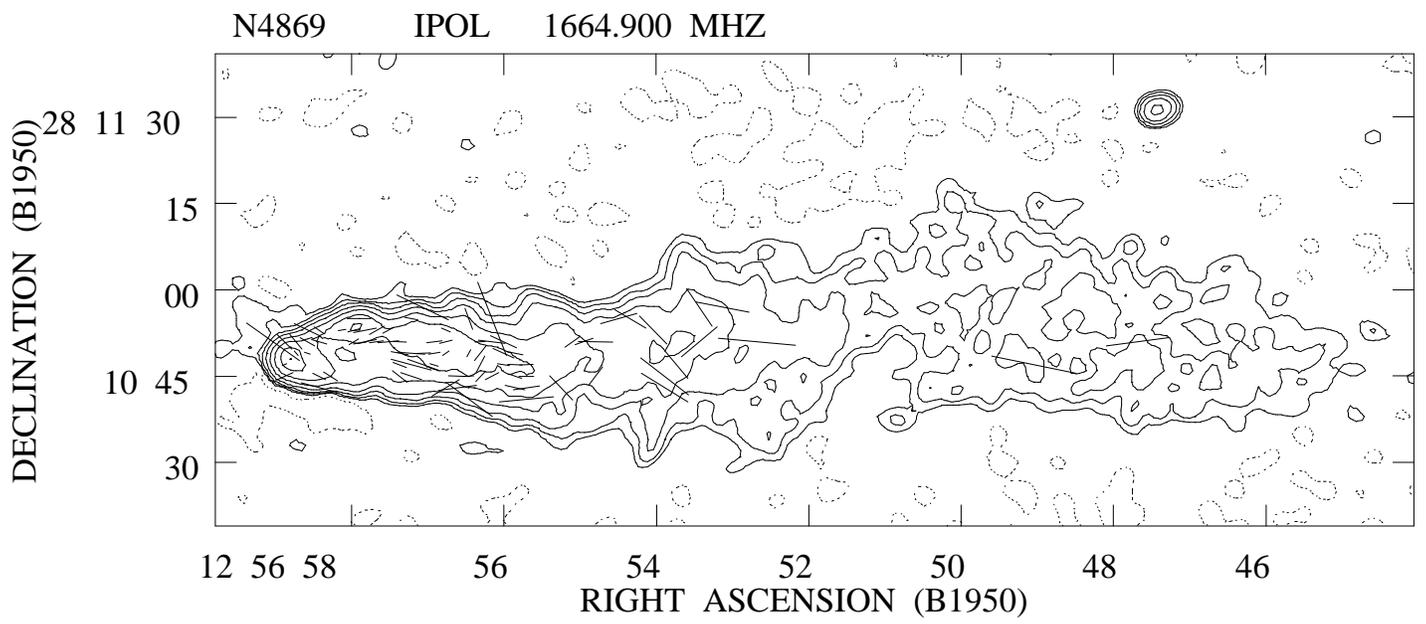
\begin{figure}
%%\special{psfile=POL6.PS.PS hoffset=-190 voffset=-535}
\picplace{1.0cm}
\caption{ Polarization image of NGC 4869 at 1665 MHz. The restoring beam is a
circular Gaussian of HPBW = 3$^{\prime \prime}$. Contour
levels of the total intensity
image are --0.12,0.12,0.2,0.35,0.7,1.5,3,6,10 mJy/beam.
The E-vector length is proportional to the polarization
percentage: 1 arcsec corresponds to 2\%.}
\end{figure}

\begin{figure}
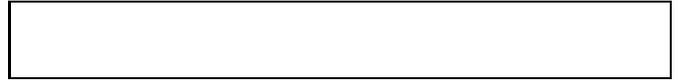

%%\special{psfile=POL6.PS.PS hoffset=-190 voffset=-535}
\picplace{1.0cm}
\caption{ Polarization image of NGC 4869 at 1465 MHz. The restoring beam is a
circular Gaussian of HPBW = 3$^{\prime \prime}$.
Contour levels of the total intensity
image are --0.09,0.09,0.18,0.35,0.7,1.5,3,6,10 mJy/beam.
The E-vector length is proportional to the polarization
percentage: 1 arcsec corresponds to 2\%.}
\end{figure}

NGC4869 is considerably polarized at 3.6 and 6 cm, all along the structure
mapped at these wavelengths, i.e. up to $\sim$80$^{\prime \prime}$ from the
core. The polarized emission is rather clumpy, with an increase of the
fractional polarization along the tail. The polarization percentage at both
frequencies is around 20-25\% in the jets, and reaches $\sim$40\% at
$\sim$70-80$^{\prime \prime}$ from the core. The polarization vector at 3.6 cm
is generally ``transversal'' ({\it perpendicular}) in the tail (Fig. 1). In the
head some structure is associated with the blobs in the jets. A significant
amount of polarized emission is associated with the short tail originating from
the S-E jet, and the local configuration of the magnetic field seems to be
circumferential (Faraday rotation at this frequency is small).

At 6 cm (Fig. 2), the fractional polarization and polarization angle are
similar to those observed at 3.6 cm, with the clear evidence of rotation of the
polarization angle in some regions.

At 18 cm, polarized emission is detected at the 5-6\% level in the head, at the
very beginning of the tail and in a few clumps further out. The highest
beginning of the tail, and in a few clumps further out. The highest fractional
polarization is of $\sim$10\% at $\sim$ 30$^{\prime \prime}$ from the core. At
further distance, where the brightness of the tail becomes lower than 0.5
mJy/beam, the detection of polarized flux is noise limited. In this case, an
upper limit of $\sim$20\% can be placed to the fractional polarization. At 20.5
cm the polarized flux is lower than at 18 cm, with peaks of 7-8\%. We note that
the fractional polarization detected at 20 cm by Kim et al. (1990) along the
tail is always below 1\%, because of their observing beam ($\sim~1^\prime$)
which is much larger than the structures in polarized emission seen in our
images, and therefore their data are severely affected by beam depolarization.

\section{Polarization comparison}

\subsection{Rotation measure}

We have derived the rotation measure (RM) of the source using the images at 3.6
and 6 cm. Using these two wavelenghts, the maximum detectable RM without
ambiguity is 1300 rad m$^{-2}$. We are confident that the RM computed with
these close wavelengths has no ambiguity. This was checked by producing two
independent maps for the two different frequencies in the C band, i.e. 4835 and
4885 MHz. Since the orientations of polarization vectors are consistent with
each other within a few degrees, the $\vert$RM$\vert$ must be lower than 500
rad m$^{-2}$.

Fig. 5 presents a grey scale image of the RM distribution. In the head region,
including the nucleus, the southern blob and its short tail, we detect the most
negative RM, with $<$RM$>$=--243 rad m$^{-2}$ and with a dispersion
$\sigma_{RM}$=87 rad m$^{-2}$. This region is about 15$^{\prime \prime}$ (5
kpc) in size, and is located at  the inner region of the optical galaxy (see
the radio-optical overlay in Dallacasa et al. 1989). The minimal values of the
RM are found in the southern blob. We note that  Kim et al. (1994) give for
this source, referring to the head, two equally likely values of RM,
--349.5$\pm$6.5 and --49.2$\pm$10.5 rad m$^{-2}$. The first value is roughly
consistent with ours, especially considering that the head region may have been
defined differently, because of the very different angular resolution.

At the beginning of the tail, i.e. immediately after the northern jet bend, the
RM increases, with an average value in the region of highest surface brightness
of $<$RM$>$=--44 rad m$^{-2}$ and with $\sigma_{RM}$=135 rad m$^{-2}$. In the
tail continuation, the average RM becomes more negative with $<$RM$>$=--197 rad
m$^{-2}$.

\begin{figure}
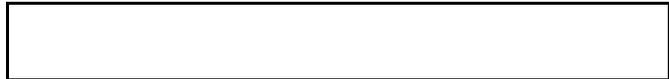

%%\special{psfile=POL6.PS.PS hoffset=-190 voffset=-535}
\picplace{1.0cm}
\caption{ Grey scale image of the RM distribution between 6 and 3.6 cm.}
\end{figure}

It is evident from Fig. 5 that the RM shows fluctuations on small scales.
Following Leahy (1984), we investigate the scale on which the fluctuations in
the RM occur using the function A($\xi$), defined as the rms difference in
rotation between all points separated by an angle $\xi$ in the map. Thus $$
A(\xi) = \sqrt {<(f(\theta) - f(\theta + \xi \phi))^2>}$$ where $\theta$ is a
position in the map and $\phi$ is a unit displacement vector; the average is
taken over all $\theta$ and all orientations of $\phi$. We note that A($\xi$)
is the square root of the structure function suggested by Simonetti et al.
(1984). For a single value of RM with random fluctuations, the rms difference
function should increase monotonically with separation until there is no
correlation, after which the value of A($\xi$) should be $\sigma\sqrt{2}$,
where $\sigma$ is the large-scale standard  deviation. The function A($\xi$)
obtained from our data is given by the error bars in Fig. 6. The dots in the
same figure represent the structure function of the RM noise map.

\begin{figure}
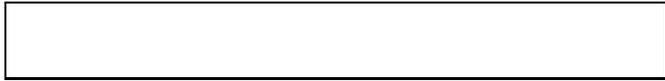

%%\special{psfile=POL6.PS.PS hoffset=-190 voffset=-535}
\picplace{1.0cm}
\caption{Structure function A($\xi$) as defined in the text. The bars refer to
the RM map, the dots to the RM noise map}
\end{figure}

The structure function is not influenced by the noise in the RM values. It
increases up to about 2.5$^{\prime \prime}$, where it shows a significant
change of slope. After this size, it flattens or increases very slowly. The
behaviour of the structure function is consistent with the existence of a
typical fluctuation size of 2.5$^{\prime\prime}$, with the possible presence of
additional fluctuations of larger scale. We note that the smallest size
obtained from the structure function corresponds to the beam size, therefore
possible smaller fluctuations would not be resolved.

The existence of small-scale fluctuations is confirmed by the behaviour of
polarization percentage in maps with increasing resolutions. In Fig. 7, we
present the profile of the fractional polarization detected at 3.6 cm along the
source tail in the map at full resolution (2.5$^{\prime\prime}$, solid line)
and in a map produced with a resolution of 4$^{\prime\prime}$ and same cellsize
(dotted line). The values of the fractional polarization in the lower
resolution are systematically lower, indicating significant beam depolarization
and confirming the existence of structure of the linearly polarized emission on
angular scales between 2.5$^{\prime\prime}$ and 4$^{\prime\prime}$

\begin{figure}
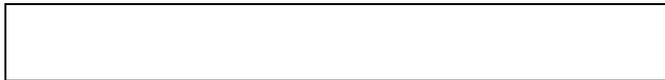

%%\special{psfile=POL7.PS.PS hoffset=-190 voffset=-535}
\picplace{1.0cm}
\caption{Fractional polarization at 3.6 cm along the ridge of the images
restored with a circular Gaussian of 2.5$^{\prime \prime}$
(solid line) and 4$^{\prime \prime}$ (dotted line).}
\end{figure}

At 18 and 20.5 cm, the orientation of the polarization angle is not consistent
with the higher frequency values and a $\lambda^2$ law. In the tail, the
polarization angle is oriented mostly longitudinally, with a few oscillations
on much larger scales than expected by the RM structure. Since the deviations
from the $\lambda^2$ law could be due to a significant effect of beam
depolarization at long wavelengths, we re-examined the 20 cm A-array data
presented by Dallacasa et al (1989), to have information with the highest
resolution. Unfortunately, no significant polarized flux was detected with the
angular resolution of 1$^{\prime \prime}$, due to the noise limitation.

\subsection{Depolarization Ratio}

We define as depolarization ratio between two wavelenghts $\lambda_1$ and
$\lambda_2$, with $\lambda_1 < \lambda_2$, the value
$$DP^{\lambda_2}_{\lambda_1}=m(\lambda_2)/m(\lambda_1)$$ where m($\lambda$)
represents the fractional polarization at the given wavelength. The
depolarization ratio between 6 and 3.6 cm, computed using only the values of
the polarization percentage  with errors $<$5\%, is around 0.8-1 (see Fig. 8),
with no evidence of any trend similar to that of the rotation measure. A plot
of the rotation measure versus the depolarization ratio (Fig. 9) confirms the
lack of any correlation between the two parameters. The source is strongly
depolarized at 18 and 20.5 cm. In the few regions where polarized flux is
detected, the depolarization ratios with respect to 3.6 cm are around 0.1-0.2.

\begin{figure}
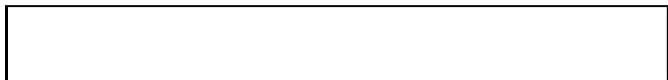

%%\special{psfile=POL8.PS.PS hoffset=-190 voffset=-535}
\picplace{1.0cm}
%% \caption{Depolarization ratio between 6 and 3.6 cm.}
\caption{Depolarization ratio between 6 and 3.6 cm, derived along the
ridge of the total intensity emission.}
\end{figure}

\begin{figure}
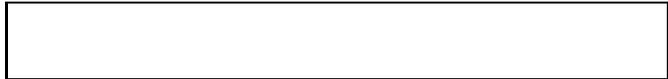

\picplace{1.0cm}
\caption{Plot of rotation measure against depolarization ratio
between 6 and 3.6 cm.}
\end{figure}

\section{Interpretation of the polarization data}

Laing (1984) warned about the difficulty of interpretation of the polarization
data, and indicated the case of deviations from $\lambda^2$ rotation associated
with depolarization as the hardest to sort out, since the Faraday effect in
this case can originate both internally within the source, and in an external
foreground screen. We summarize in the following the relevant arguments useful
for the interpretation of polarization data.

\subsection{Internal Faraday effect}

In case that the Faraday effect originates entirely within the source, Burn
(1966) predicts that the polarization angle will obey a $\lambda^2$ law of
rotation and that significant rotation will be accompanied by severe
depolarization. Assuming a simple source geometry (slab), the polarization
percentage at the wavelength $\lambda$, P$_{\lambda}$, is reduced with respect
to the intrinsic polarization P$_i$ according to the formula $$P_{\lambda} =
P_i{{sin (RM\lambda^2)}\over{RM \lambda^2}} \eqno(1) $$ while the positon angle
$\chi$ rotates as $\lambda^2$ in the range 0$\le \chi \le \pi/2$, and then
changes regularly showing  a "sawtooth" variation (Laing 1984). In more
realistic mixed geometries, P$_{\lambda}$ does not have zeroes and the
polarization angle will obey a $\lambda^2$ rotation over at most 90$^{\circ}$.

\subsection{External uniform screen}

In the simplest case of Faraday depth effectively constant across the beam, the
Faraday effect from a screen (single slab) external to the source and with
uniform magnetic field produces no depolarization and a rotation of the
polarization angle given by $\lambda^2<$RM$>$, with $$ <RM> = 812 B_{\parallel}
n_e d  \eqno(2) $$ where n$_e$ is the electron density in cm$^{-3}$,
B$_{\parallel}$ is the strength of magnetic field along the line of sight in
$\mu$G, and $d$ is the depth of the screen in kpc.

\subsection{External screen with tangled magnetic field}

The existence of small scale structure of the magnetic field in an external
screen is relevant to produce both rotation of the polarization angle and
depolarization. The effect of a Faraday screen with a tangled magnetic field
has been analyzed by Lawler and Dennison (1982) and by Tribble (1991). In the
simplest ideal case, the screen is made of cells of uniform size, electron
density and magnetic field strength, but with field orientation at random
angles in each cell. The observed RM along any given line of sight will be
generated by a random walk process, and the distribution of the RM results a
gaussian with zero mean and variance given by $$ \sigma_{RM}= {{812} \over
{\sqrt 3}} n_eBN^{1/2}l \eqno(3) $$ where $n_e$ is the electron density in
cm$^{-3}$, B is the magnetic field strength in $\mu$G, N is the number of cells
along the line of sight and $l$ is the size of each cell in kpc. The
depolarization produced by a Faraday screen with tangled magnetic field can be
approximated at a long wavelength $\lambda$ by the formula suggested by Tribble
(1991), which includes dependence on the uniform cell size and RM dispersion $$
DP \sim {{\theta}\over{\theta_b}}\cdot{0.6\over {\sigma_{RM} \lambda^2}}
\eqno(4) $$ where $\theta$ is the angular size of each cell and $\theta_b$ is
the size of the observing beam.

\subsection{External screen with tangled magnetic field and gas
density distribution}

In the case of a tangled magnetic field,  the value of $\sigma_{RM}$ can also
be obtained for a more realistic gas density distribution. It has been found
that the gas density in clusters of galaxies follows a hydrostatic isothermal
beta model (Cavaliere and  Fusco-Femiano 1981), i.e. $$
n_e(r)=n_0(1+r^2/r_c^2)^{-3\beta/2} \eqno (5)$$ where $n_0$ is the central
electron density, and $r_c$ is the core radius of the gas distribution.  In
this case, Kim et al. (1991) found $$ \sigma_{RM}= {{688 B}\over
{(1+r^2/r_c^2)^{(6\beta -1)/4}}} {{\Gamma(3\beta-0.5)}\over{\Gamma(3\beta)}}
 n_0  M^{1/2}l \eqno(6) $$ where M is the number of cells per core radius. For
$\beta$=0.7 the previous formula becomes $$ \sigma_{RM} \approx {{585 B}\over
{(1+r^2/r_c^2)^{0.8}}}  n_0  M^{1/2}l. \eqno(7) $$

\section{Discussion}

The observational properties of NGC4869 derived from the polarization data can
be summarized as follows: \begin{itemize} \begin{enumerate} \item the RM is
characterized by the presence of local fluctuations, occurring on typical
scales of $\sim$2.5$^{\prime \prime}$. Moreover, it is remarkable that the RM
in the head of the source is more negative than in the other regions, with the
lowest values in the southern blob;

\item depolarization ratios at 6 cm with respect to 3.6 cm are
between 0.8 and 1, while at at 18 and 20.5 cm the source is generally
much strongly depolarized;

\item the
depolarization ratio seem not to be correlated with RM;

\item  the orientation of the polarization angle at 18 and 20.5 cm, in the
regions where significant polarized flux is detected, shows deviation from a
$\lambda^2$ law, defined using the angles observed at 3.6 and 6 cm.
\end{enumerate}
\end{itemize}

We note that the Coma cluster lies close to the north galactic pole, therefore
the effect of the interstellar medium of the Milky Way Galaxy on the
polarization is negligible.

In NGC4869 the RM distribution is not correlated with the depolarization ratio
and this favours the idea that the Faraday effect is mostly external to the
source. In this case the deviation from the $\lambda^2$ law at long wavelength
would be due to beam depolarization. This is consistent with the existence of
irregularities in the foreground screen smaller than the observing beam at 20.5
cm.

\subsection{Effect of the galaxy NGC~4869}

The RM in the head of the source could be affected by the interstellar medium
of the galaxy NGC~4869, as already found in the radio galaxy NGC~6047 (Feretti
\& Giovannini 1988).

However, the scale of the RM fluctuations in the head is not different from
that in the tail and also the depolarization of the head is very similar to
that in the other source regions. Therefore, we do not have a definite evidence
of the existence of magnetic field associated with the interstellar medium of
the galaxy. In any case, since this magnetic field does not seem to affect the
dispersion of the RM values, it shoud be ordered on scales comparable to the
galaxy size.

We note that the southern blob in NGC~4869 exhibits a more negative RM than the
northern one, and this could arise from the jet orientation. The northern jet
is approaching the observer, while the southern one is receding from us
(Feretti et al. 1990).

\subsection{Effect of the halo Coma C with an ordered magnetic field}

The existence of magnetic field associated with the Coma intergalactic medium
is directly deduced from the presence of the the diffuse radio halo Coma C,
which is permeating the central region of the Coma cluster, for a total size of
25' (Giovannini et al. 1993), corresponding to $\sim$500 kpc. No polarization
data of Coma C are available to get direct information on the magnetic field
structure. The equipartition magnetic field of the halo is 0.5 $\mu$G
(Giovannini et al. 1993). According to X-ray data (Hughes 1988), the gas
density distribution follows an hydrostatic isothermal model with central gas
density  $n_0$ = 3.7$\times$10$^{-3}$ cm$^{-3}$, core radius = 9.8'
(corresponding to 198 kpc) and $\beta$ = 0.76. According to the orbit  computed
by Feretti et al. (1990), NGC 4869 is imbedded within the halo, and not simply
projected onto it. Assuming equipartition conditions for the halo, i.e. a
magnetic field along the line of sight = 0.5 $\mu$G, and that the magnetic
field is completely ordered, a rotation measure $\vert$RM$\vert$ $\approx$ 300
rad/m$^2$ is implied (equation 2), constant across the source. While the value
of the average RM can be different, depending  on the Faraday depth through the
cluster, the implication of a constant value across the  source is not
consistent with our data, given the large detected dispersion in the values of
RM.

\subsection{Effect of a cluster tangled magnetic field}

The dispersion of RM observed across NGC 4869 can be explained by assuming that
the magnetic field is tangled on typical scales of the same size as the RM
fluctuations, i.e. 2.5$^{\prime\prime}$, corresponding to 0.85 kpc. This leads
to $\sim$235 cells per cluster core radius. Moreover, field reversals must take
place. The RM distribution, obtained using only the values of the polarization
angle with errors $<$10$^{\circ}$, gives $<$RM$>$ = --127 rad m$^{-2}$, with a
dispersion $\sigma_{RM}$=181 rad m$^{-2}$. Using the equation (6), assuming a
projected distance of 5$^{\prime}$ from the cluster center, we obtain a
magnetic field associated with  the intergalactic medium B = 8.5
$h^{1/2}_{100}~~\mu$G, where $h_{100}$= H$_0$/100 km s$^{-1}$ Mpc$^{-1}$. This
value refers to  the trivial case that the projected distance from the cluster
center coincides with the true distance. Projection effects obviously may play
an important role. A displacement of NGC4869 from the cluster center also along
the line of sight  by about   $\pm$3.5$^{\prime}$ (see Feretti et al. 1990),
leads to an uncertainty of about $\pm$1.5 $\mu$G on the value of the magnetic
field.

With the assumptions of a tangled magnetic field, the expected depolarization
ratios at 18 and 20.5 cm, estimated with equation (4), are \ltsim 0.08-0.06,
roughly consistent with the observational result.

However, a Faraday screen with a tangled magnetic field should produce, as
noted in \S 5.3, a RM distribution with zero mean, which is not the case.
Therefore, we have to assume a more complex scenario to explain our data.

\subsection{A tangled magnetic field plus a uniform disk}

The non zero average RM could be due to the existence of a weak magnetic field
component, ordered on a large scale, as suggested by Taylor and Perley (1993)
for the source Hydra A. The average RM in the tail  is $<$RM$>$ =  --127 rad
m$^{-2}$, which leads to a magnetic field component ordered on a scale of
$\sim$200 kpc, with strength of B=0.1-0.3~$\mu$G, depending on the location of
the galaxy along the line of sight.

Kim et al. (1990) and Kim et al. (1994), in a survey of the Coma cluster,
obtained Faraday rotation measures of 18 sources at different projected
distances from the cluster center and found a significant contribution to the
RM of sources seen through the cluster. The sources within 15$^{\prime}$ from
the cluster center, besides our source, are: 5C4.70, 5C4.74 and 5C4.85, which
have RM of 47$\pm$7, --65$\pm$7 and --32$\pm$10 rad m$^{-2}$, respectively.
These values of RM are significantly different from the average RM of the
present source NGC4869. We note, however, that 5C4.70 and 5C.74 are at more
than one core radius from the cluster center, therefore we must consider
different Faraday depths, deriving from lower electron densities. The
discrepancy between the the average RM of our source, and the value of the RM
of 5C4.85, identified with the dominant galaxy of the Coma cluster, could arise
from projection effects, and would therefore imply that NGC4869 is actually
located beyond NGC4874. Moreover, a contribution from internal origin to the RM
of NGC4874 could be present. New high resolution and sensitivity observations
would be  needed to clarify this point.

\subsection{The Coma magnetic field and implications on
the conditions of the halo Coma C}

The RM value and distribution of NGC4869 can be reasonably explained assuming a
two-component magnetic field, one tangled on the kpc scale, the other organized
over a scale of 200 kpc. The tangled magnetic field has a strength of about
8.5$\pm$1.5 $\mu$G, while the uniform component is much weaker, about
0.2$\pm$0.1 $\mu$G.

A magnetic field of several $\mu$G, associated with the intergalactic medium of
clusters of galaxies, is not surprising. The existence of cluster magnetic
field is discussed in many recent papers (Ge \& Owen 1993, Taylor \& Perley
1993, Taylor et al. 1994, Ge \& Owen 1994). Generally these cases refer to
clusters with cooling flow, where the magnetic field is suggested to be
amplified by compression and inflow (Soker \& Sarazin 1990). The strength of
the magnetic field has been observed to be up to $\sim$30 $\mu$G. The Coma
cluster is the first case of  a cluster without the cooling flow, where the
existence of a magnetic field of several $\mu$G is suggested.

Even allowing for the uncertainties related with projection effects, the Coma
magnetic field derived in Sect. 6.3 is more than one order of magnitude larger
than the value (B$_{eq}$ = 0.5$\mu$G) derived by Giovannini et al. (1993), by
integrating the total luminosity of the source between 10 MHz and 1 GHz, with a
spectral index $\alpha$=0.5 (the low frequency one), assuming energy
equipartition between magnetic field and relativistic particles, a filling
factor $\Phi$ = 1 and equal energy in electrons and heavy particles (k=1). We
note that the ratio between the magnetic field derived from RM arguments, and
the minimum equipartition value is weakly affected by the assumed value of the
Hubble constant. Infact, since the equipartition magnetic field scales as
$H^{2/7}_{0}$, the ratio B/B$_{eq}$ depends on H$^{3/14}_{0}$ (i.e. the ratio
only decreases from 17 to 15 if H$_0$=50 is used instead of H$_0$=100).

The large difference between the magnetic field estimated here and B$_{eq}$
could be explained, without abandoning the equipartition hypothesis, assuming a
low filling factor ($\Phi <$ 1) and a large ratio between the proton and
electron energy (k$>$ 1). Possible values  of the two parameters, whose
combination raises the equipartion magnetic field to 8.5 $\mu$G are given in
Table 2.
\begin{table}
\caption{Values of the proton to electron energy ratio k, and of the
filling factor $\Phi$ needed to raise the equipartition magnetic field
to 8.5$ \mu$G. }
\begin{flushleft}
\begin{tabular}{llclcll}
\hline
\noalign{\smallskip}
{}~&~&  k  &~& $\phi$ ~&~&\\
\noalign{\smallskip}
\hline
\noalign{\smallskip}
{}~&~&   1                &~&       5   $\times 10^{-5}$  ~&~& \\
{}~&~&   10               &~&       2.7 $\times 10^{-4}$  ~&~& \\
{}~&~&   $ 10^2$          &~&       2.5 $\times 10^{-3}$  ~&~& \\
%~&~&   2 $\times 10^2$  &~&       5   $\times 10^{-3}$  ~&~& \\
{}~&~&   4 $\times 10^2$  &~&       1   $\times 10^{-2}$  ~&~& \\
{}~&~&   1 $\times 10^3$  &~&       2.5 $\times 10^{-2}$  ~&~& \\
%~&~&   2 $\times 10^3$  &~&       5   $\times 10^{-2}$  ~&~& \\
{}~&~&   4 $\times 10^3$  &~&       0.1	    	      ~&~& \\
{}~&~&   1 $\times 10^4$  &~&       0.25	     	      ~&~& \\
{}~&~&   4 $\times 10^4$  &~&       1	              ~&~& \\
\noalign{\smallskip}
\hline
\noalign{\smallskip}
\end{tabular}
\end{flushleft}
\end{table}

We derive that values of k between 10 and 100 (see Feretti et al. 1992 and
references therein) imply very low filling factors. Therefore, we favour the
possibility that Coma C is not at the equipartition.

In any case, the total energy content in Coma C is not minimum, but at least
100 times larger than the minimum equipartition value. This would in turn
produce a larger internal pressure, thus partly reducing the apparent imbalance
between internal non-thermal pressure and external pressure of the X-Ray gas.

The high energy content of the radio halo Coma C has important implications on
the physics of this puzzling source. Owing to the  large value of the magnetic
field, the time for synchrotron radiation losses of electrons in the halo is
reduced by at least one order of magnitude, thus reinforcing the need for
reacceleration processes. The energy required to sustain the radio emission
from the halo is very large and the contribution of galactic wakes in supplying
energy from turbolence is not enough to support the existence of Coma C (see
Giovannini et al. 1993). From the X-ray map (Briel et al., 1992), it is evident
that the Coma cluster is not yet relaxed.Therefore the energy available from
the merger process (Tribble 1993) of the NGC~4839 sub-group (Briel et al.,
1992) into the main cluster becomes very important for the energy supply of
Coma C (Giovannini et al., 1993, Burns et al., 1994). Since thermal and non
thermal particles are mixed inside Coma C, we can reasonably assume that for
the same reason the Coma magnetic field and relativistic particles are not yet
at the equilibrium (equipartition energy condition).

\section{Conclusions}

The main conclusions of this paper are the following:
\begin{itemize}
\begin{enumerate}
\item The large dispersion of the RM distribution of the source
NGC~4869 can be interpreted as originating from a cluster magnetic
field of 8.5 $\mu$G, tangled on scales of less than 1 kpc.  Large
magnetic fields have been so far found in clusters showing a cooling
flow. This is the first case of a large magnetic field associated with
a cluster without a cooling flow.

\item The non zero average of the RM suggests the presence of a weaker
magnetic field component, of 0.2 $\mu$G, uniform on a cluster core
radius scale ($\sim$ 200 kpc).

\item The value of the magnetic field derived in the Coma cluster is
more than one order of magnitude larger than the minimum equipartition
value computed for the halo source Coma C. This implies that the radio
halo is likely to be out of energy equipartition and
reinforces the need for reacceleration processes for the relativistic
electrons radiating in Coma C. The energy supply in Coma C is likely
to be the merger process.
\end{enumerate}
\end{itemize}

\begin{acknowledgements}
We thank R. Fanti for useful suggestions and discussions and M.
Bondi for his help with the structure function. We are grateful to
P. Kronberg, the referee, for helpful comments.
The National Radio Astronomy Observatory (NRAO) is operated by Associated
Univesities, Inc., under contract with the National Science Foundation.
\end{acknowledgements}

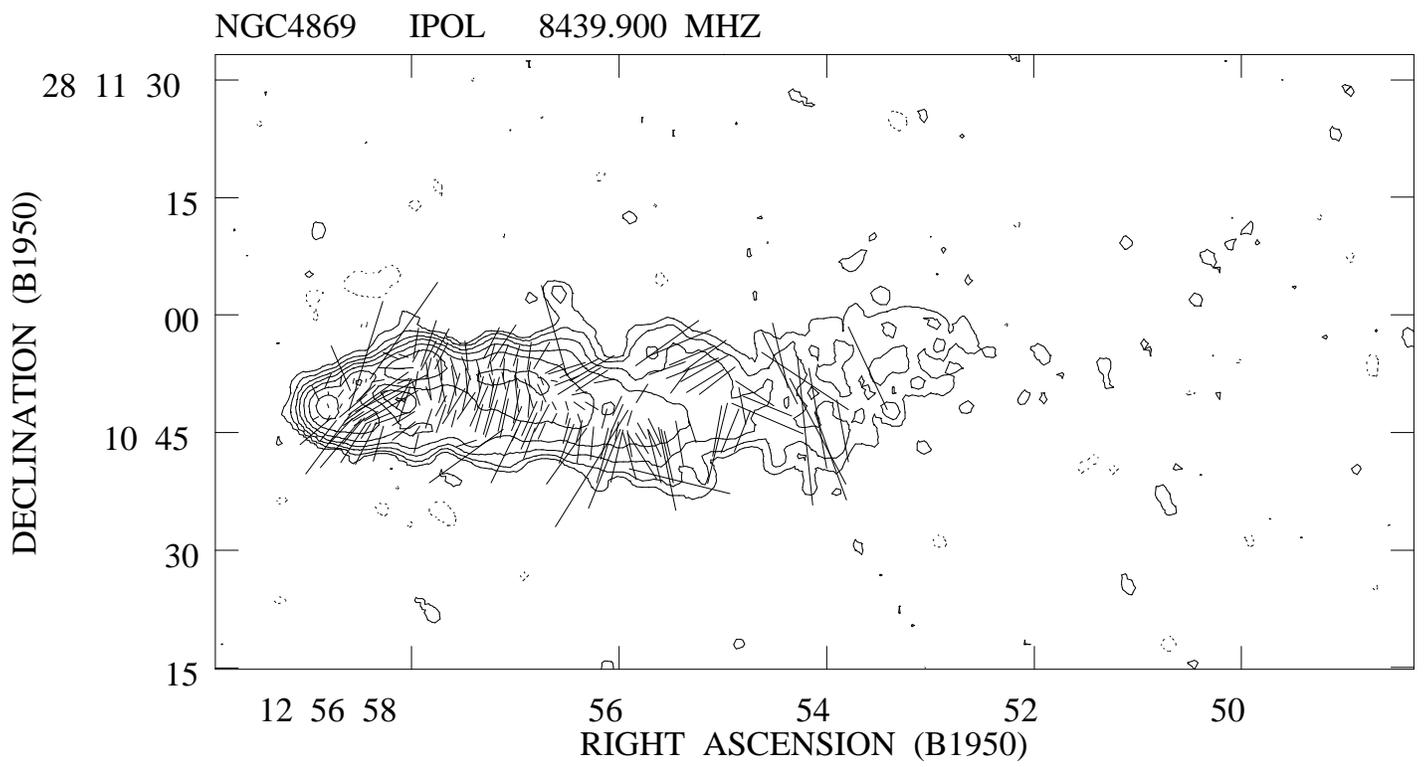
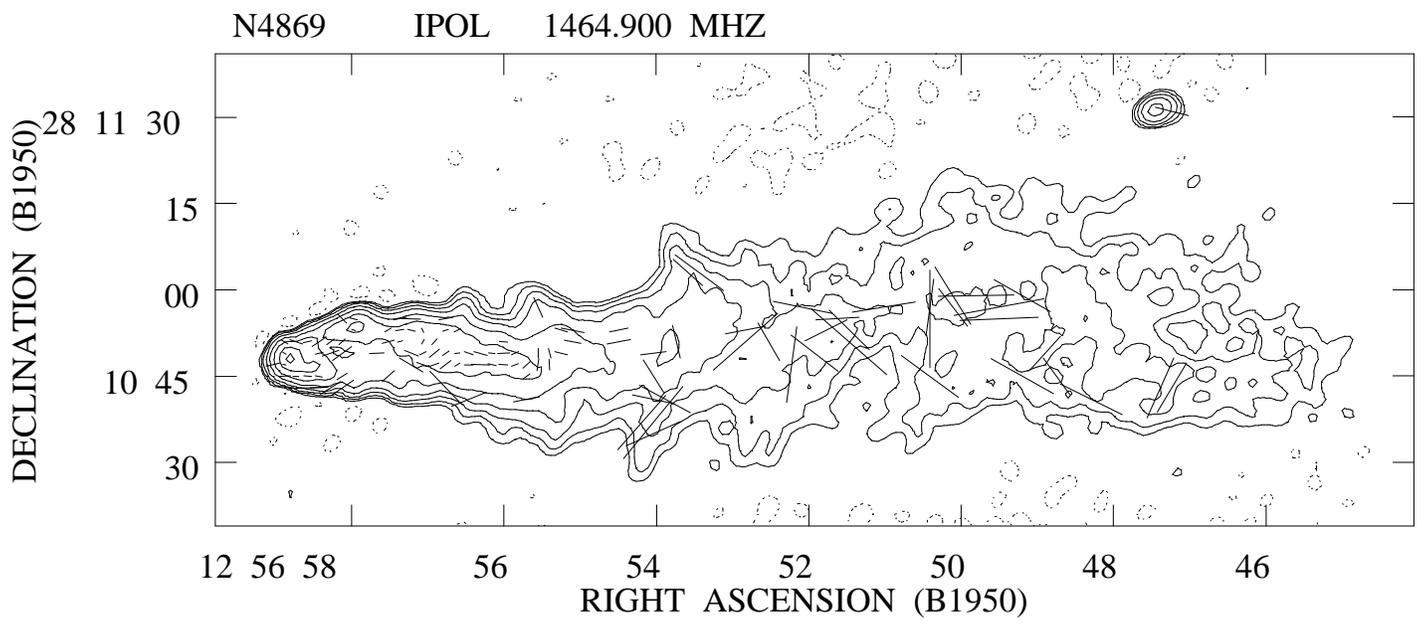
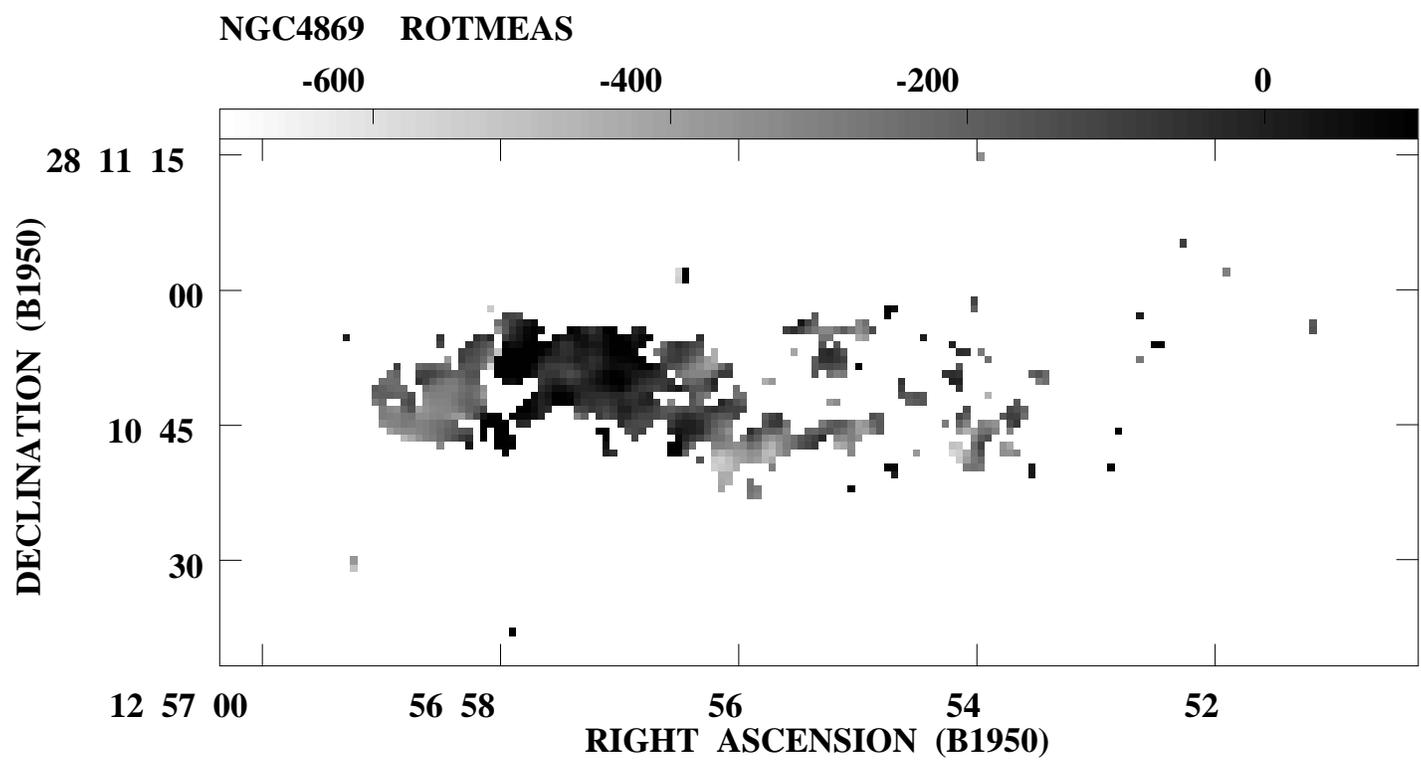
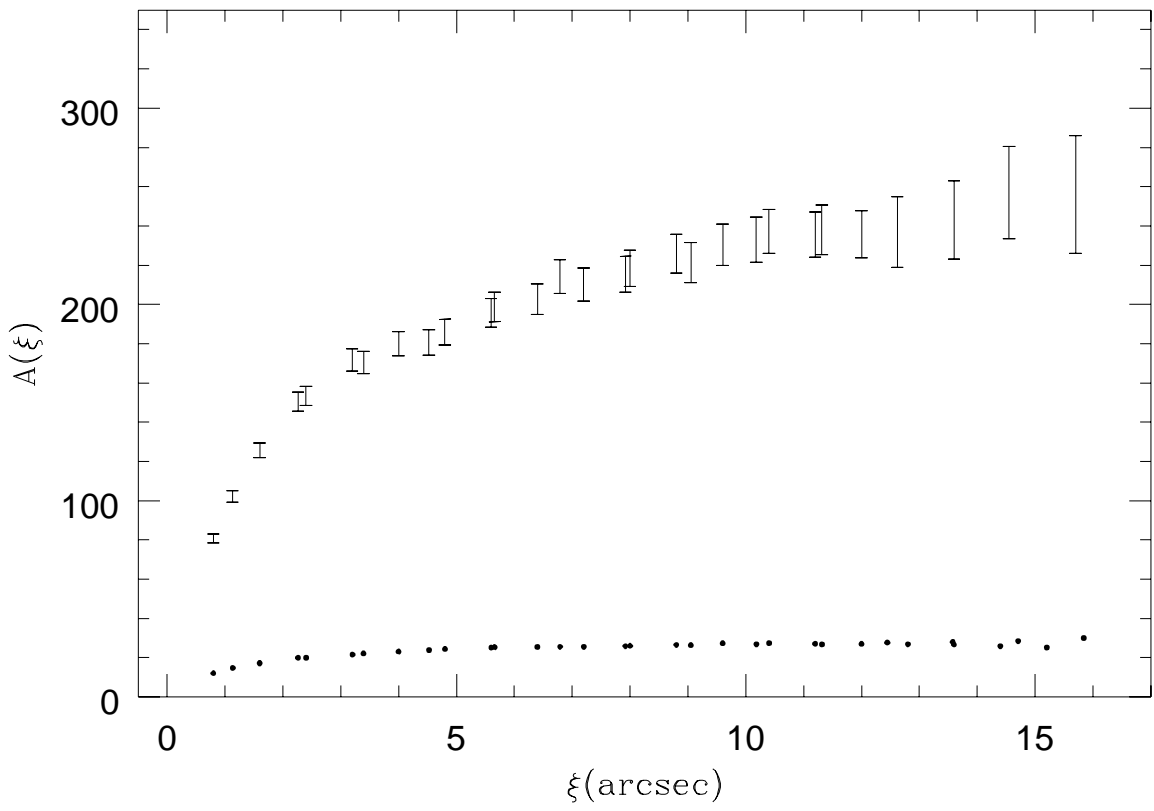
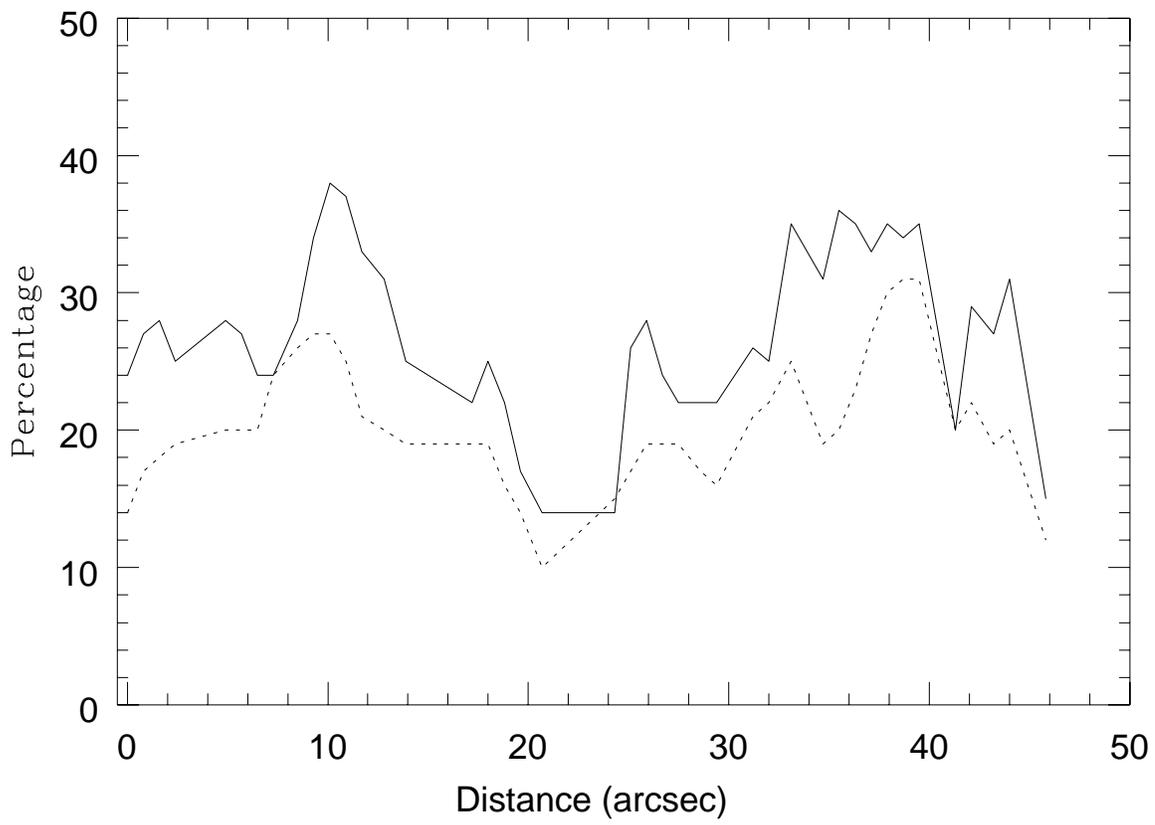
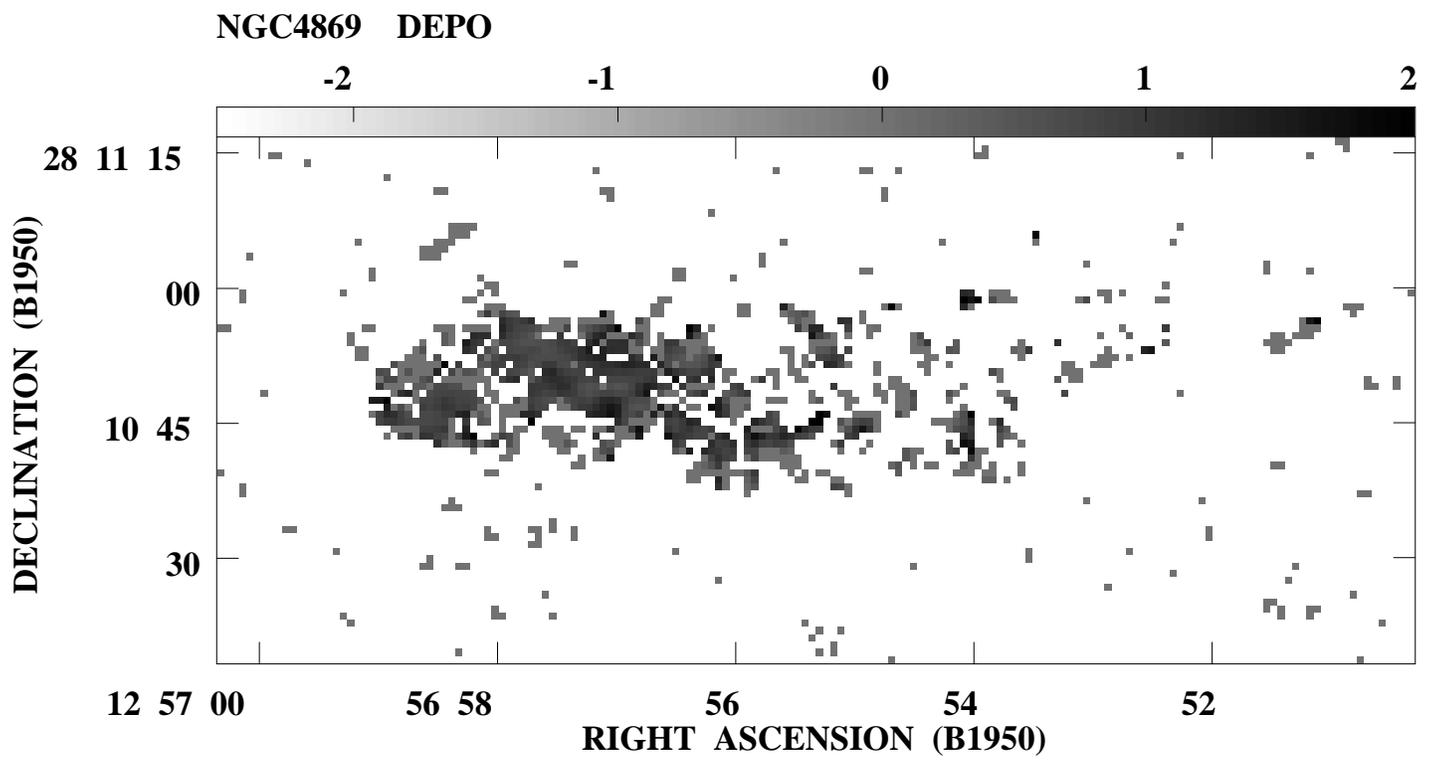
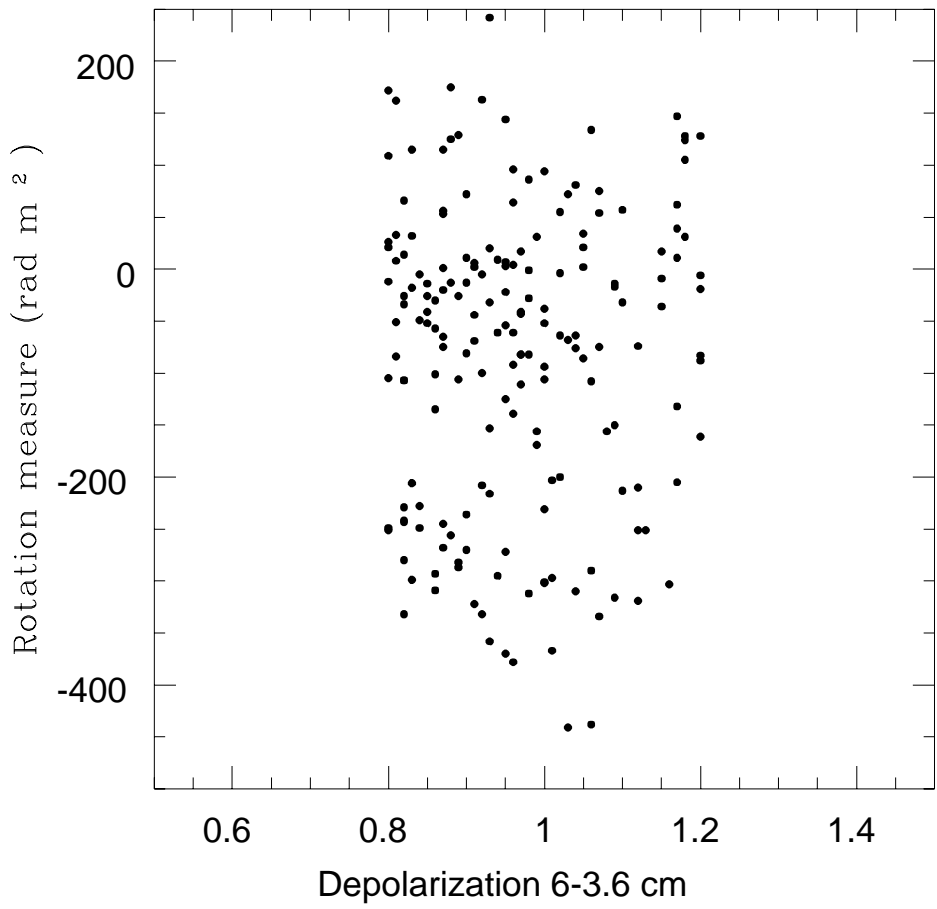
\end{document}